\documentclass[preprint,aps,10pt]{revtex4}
\usepackage{graphicx}
\def\beq{\begin{equation}}
\def\eeq{\end{equation}}
\def\bea{\begin{eqnarray}}
\def\eea{\end{eqnarray}}

\def\as{\alpha_s}

\newcommand{\bsgam}{\mbox{$B \to X_s \gamma\,$}}

\newcommand{\bsll}{\mbox{$B\rightarrow X_s \ell^+ \ell^-$}}

\def\sh{\hat{s}}
\def \branch{{\cal B}}
\def \eff{\hbox{eff}}
\def \Im{{\hbox{Im}}\,}
\def \Re{{\hbox{Re}}\,}
\def \gev{{\hbox{GeV}}}

\def \nn{\nonumber}
\newcommand{\pole}{\text{pole}}

\begin{document}
\title{ A Simple Approach to Fourth Generation Effects in $B\rightarrow X_s \ell^+ \ell^-$ Decay}
\author{Levent Solmaz}
 \email{lsolmaz@photon.physics.metu.edu.tr,lsolmaz@balikesir.edu.tr}
\affiliation{
Balikesir University, Physics Department (32), Balikesir, Turkey}

\date{\today}
\begin{abstract}
  In a scenario in which fourth generation fermions exist, we study effects
  of new physics on the differential decay width,
  forward-backward asymmetry 
  and integrated branching ratio for $B\rightarrow X_s \ell^+ \ell^-$ decay with $(\ell=e,\mu)$. Prediction of the new
  physics on the mentioned quantities essentially differs from the Standard Model
  results, in certain regions of the parameter space, enhancement of new physics on the above mentioned physical quantities
  can yield values as large as two times of the SM predictions, whence
  present limits of experimental measurements of branching ratio is spanned, contraints of the new physics can be extracted.
  For the fourth generation CKM factor $V_{t^\prime b}^\ast V_{t^\prime
  s}$ we use $\pm 10^{-2}$ and $\pm 10^{-3}$ ranges, take into consideration the possibility of a complex phase
  where it may  bring sizable contributions, obtained no significant dependency on the imaginary part of the new CKM factor.
  For the above mentioned quantities with a new family, deviations from the
  SM are promising, can be used as a probe of new physics.
\end{abstract}
\maketitle

\section{Introduction}
Even if  Standard Model (SM) is a  successful theory, one should
also check probable effects that  may come from potential new
physics. In the SM, since we do not have a clear theoretical
argument to restrict number of generations to three, possibility
of a new generation should not be ruled out until there is a
certain evidence which order us to do so. This is especially true
for rare B decays, which are very sensitive to generic expansions
of the SM, due to their loop structure.
 We know from neutrino experiments that, for the mass of
the extra generations there is a lower bound for the new
generations ($m_{\nu_4}> 45~ GeV$) \cite{Ref1}. Probable effects
of extra generations  was studied in many works
\cite{Ref2}--\cite{Ref16}. The existing electroweak data on the
$Z$--boson parameters, the $W$ boson and the top quark masses
excluded the existence of the new generations with all fermions
heavier than the $Z$ boson mass \cite{Ref16}, nevertheless, the
same data allows a few extra generations, if one allows neutral
leptons to have masses close to $50~GeV$. In addition to this,
recently observed neutrino oscillations requires an enlarged
neutrino sector \cite{neutrino}.

Generalizations of the SM can be used to introduce  a new family,
which was performed previously \cite{Ref17}. Using similar
techniques, one can search fourth generation effects in  B  meson
decays. The contributions from fourth generation to rare decays
have been extensively
studied~\cite{Hou,wshou,fourth,Huo,aliev2003}, where the measured
decay rate has been used to put stringent constraints on the
additional CKM matrix elements. In addition to \bsgam, \bsll~ can
be mentioned as one of the most promising areas in search of the
fourth generation, via its indirect loop effects, to constrain
$V_{t^\prime b}^\ast V_{t^\prime s}$ \cite{Ref21,Ref210}. The
restrictions of the parameter space of nonstandard models based
on LO analysis  are not as sensitive as in the case of NLO
analysis, hence a NLO analysis considering the possibility of a
complex phase is important, which we plan to revise \cite{mywork}.

 On
the experimental side, the inclusive $B \to X_s \ell^+ \ell^-$
(with $\sqrt{q^2} > 0.2$ GeV)  decay with electron and muon modes
combined ($\ell=e,\mu$) have been observed (Belle
\cite{Kaneko:2002mr}), (BaBar \cite{Aubert:2003rv}),
\begin{eqnarray}
{\mathcal{B}}(B \to X_s \ell^+ \ell^-) & = & (6.1 \pm
1.4^{+1.4}_{-1.1}) \cdot 10^{-6} ,\;\;\;\;\;\;\;\; \\
{\mathcal{B}}(B \to X_s \ell^+ \ell^-) & = & (6.3 \pm
1.6^{+1.8}_{-1.5}) \cdot 10^{-6} \; .
\end{eqnarray}

They are in agreement with the SM ${\mathcal{B}}(B \to X_s \ell^+
\ell^-)_{SM}=4.2 \pm 0.7 \cdot 10^{-6} $ for the same cuts
\cite{Ali:2002jg}.

 On the theoretical side, situation within  and beyond the SM is
well settled. A collective theoretical effort  has led to the
practical determination of $\bsll$ at the NNLO,  which  was
completed recently, as a joint effort of different groups
(\cite{BMU,Asatrian:2002va,Ghinculov:2002pe}), and references
therein. It is necessary to have precise calculations also in the
extensions of the SM, which was performed for certain models. With
the appearance  of more accurate data we might be able to provide
stringent constrains on the free parameters of the models beyond
SM. From this respect, a NNLO analysis of the new generation is
important. We study the contribution of the fourth generation in
the rare \bsll decay at NNLO, to obtain experimentally measurable
quantities which is expected to appear in the forthcoming years.

The paper is organized as follows. In section 2, we present the
necessary theoretical expressions for the \bsll  decay in the SM
with four generations. Section 3 is devoted to our conclusion.
\section{\bsll~decay and fourth generation}
We use the framework of an  effective low-energy theory, obtained
by  integrating  out heavy degrees of freedoms, which in our case
W-boson and top quark and an additional $t^{\prime}$ quark. Mass
of the $t^{\prime}$ is at the order of $\mu_W$. In this
approximation the effective Hamiltonian relevant for the \bsll
decay reads \cite{AAGW}
\begin{eqnarray}
    \label{Heff}
    {\cal H}_{\eff} =  - \frac{4G_F}{\sqrt{2}} V_{ts}^* V_{tb}
    \sum_{i=1}^{10} C_i(\mu) \, O_i (\mu)\quad ,
\end{eqnarray}

where $G_F$ is the Fermi coupling constant  $V$ is the
Cabibbo-Kobayashi-Maskawa (CKM) quark mixing matrix, the the full
set of the operators ${O}_i(\mu)$ and the corresponding
expressions for the Wilson coefficients ${C}_i(\mu)$ in the SM
can be found in Ref.\cite{BMU}.

 In the model under consideration, the fourth generation is
introduced in a similar  way the three generations are introduced
in the SM, no new operators appear and clearly the full operator
set is exactly the same as in SM, which is a rough approximation.
The fourth generation changes values of the Wilson coefficients
$C_i(\mu)$, $i=7,8,9$ and $10$, via virtual exchange of the
fourth generation up quark $t^\prime$. With the definitions
$\lambda_{j}=V_{j s}^{\ast} V_{j b}\text{,
where}~j={u,~c,~t,~t^\prime}$,
  the new physics Wilson coefficients can be written in the following
form
\beq
 C_i^{4G}(\mu_W) =  \frac{\lambda_{t^\prime}}{\lambda_{t}} C_i(\mu_W)_{m_t\rightarrow m_{t^\prime}} ~,
 \eeq
where the last terms in these expression describes the
contributions of the $t^\prime$ quark to the Wilson coefficients
with the replacement  of $m_t$ with $m_t\prime$. Notice that we
use the definition $\lambda_{t^\prime}=$ $V_{t^\prime s}^{\ast}
V_{t^\prime b}$ which  is the element of the $4\times 4$
Cabibbo--Kobayashi--Maskawa (CKM) matrix, from now on '4G' will
stand for sequential fourth generation model. In this model
properties of the new $t^\prime$ quark are the same as ordinary
$t$, except its mass and corresponding CKM couplings. A few
comments are in order here: to obtain quantitative results we
need the value of the fourth generation CKM matrix element $
V_{t^\prime s}^\ast V_{tb}$ which can be extracted i.e. from
\bsgam~decay as a function of mass of the new top quark
$m_t^{\prime}$. For this aim following \cite{Ref21,Ref210}, we
can use the fourth generation CKM factor $\lambda_{t^{\prime}}$
in the range $-10^{-2}\leq\lambda_{t^{\prime}}\leq 10^{-2}$. In
the numerical analysis, as a first step, ${\lambda_{t^{\prime}}}$
is assumed real and expressions are obtained as a function of
mass of the extra generation top quark $m_{t^{\prime}}$. It is
interesting to notice that, if we assume
 $\lambda_{t^{\prime}}$ can have imaginary parts, experimental values can  also be
 satisfied \cite{mywork,aliev2003}. Nevertheless, if we impose the unitarity condition of the CKM matrix we have
 \bea V_{us}^\ast V_{ub} + V_{cs}^\ast V_{cb} + V_{ts}^\ast
V_{tb} + V_{t^\prime s}^\ast V_{t^\prime b} = 0 ~.
\label{unity}\eea With the values of the CKM matrix elements in
the SM \cite{Ref29}, the sum of the first three terms in Eq. (5)
is about $7.6 \times 10^{-2}$, where the error in sum of first
three terms  is about $\pm 0.6 \times 10^{-2}$. We assume the
value of $\lambda_{t^\prime}$ is within this error range.

 What should not be ignored  in constraining $\lambda_{t^\prime}$ is that, when  adding a fourth family
the present constrains on the elements of CKM  may get relaxed
\cite{Ref29}. In order to have a clear picture of
$\lambda_{t^\prime}$, CKM matrix elements should be calculated
with the possibility of a new family, using present experiments
that constitutes the CKM. From this respect we do not have to
exclude certain regions that violate unitarity of the present CKM,
but take it in the ranges $-10^{-2}\leq\lambda_{t^{\prime}}\leq
10^{-2}$ and $-10^{-3}\leq\lambda_{t^{\prime}}\leq 10^{-3}$.

\subsection{Differential Decay Width}
Since extended models are very sensitive to NNLO corrections, we
used the NNLO expression for the branching ratio of the radiative
decay \bsll, which has been presented in Refs.
\cite{AAGW,Ali:2002jg}. In the NNLO approximation, the invariant
dilepton mass distribution for the inclusive decay $B \to X_s
\ell^+ \ell^-$ can be written as
\begin{eqnarray}
\label{rarewidth}
    \frac{d\Gamma(b\to X_s \ell^+\ell^-)}{d\hat s}&=&
    \left(\frac{\alpha_{em}}{4\pi}\right)^2
    \frac{G_F^2 m_{b,pole}^5\left|V_{ts}^*V_{tb}^{}\right|^2}
    {48\pi^3}(1-\hat s)^2 \nn \\
    &&\times\left ( \left (1+2\hat s\right)
    \left (\left |\widetilde C_9^{\eff}\right |^2+
    \left |\widetilde C_{10}^{\eff}\right |^2 \right )
    + 4\left(1+2/\hat s\right)\left
    |\widetilde C_7^{\eff}\right |^2+
    12 \mbox{Re}\left (\widetilde C_7^{\eff}
    \widetilde C_9^{\eff*}\right ) \right ) \, ,
\end{eqnarray}
where $\hat s=m_{\ell^{+}\ell^{-}}^2/m_{b,pole}^2$ with ($\ell=e
\text{~or~}\mu $). In the SM the effective Wilson coefficients
$\tilde{C}_7^{\eff}$, $\tilde{C}_9^{\eff}$ and
$\tilde{C}_{10}^{\eff}$ are given by~\cite{BMU,AAGW} and can be
obtained from Eqs.(\ref{effcoeff7new},\ref{effcoeff9new} and
\ref{effcoeff10new}), by setting $4G\rightarrow~0$. Following the
lines of A.Ali \cite{Ali:2002jg} with the assumption that only
the lowest non-trivial order of these Wilson coefficients get
modified by new physics, which means that $C_7^{(1)}(\mu_W)$,
$C_8^{(1)}(\mu_W)$, $C_9^{(1)}(\mu_W)$ and  $C_{10}^{(1)}(\mu_W)$
get modified, the shifts of the Wilson coefficients at $\mu_W$
can be written  as
\begin{equation}
C_i(\mu_W) \longrightarrow C_i(\mu_W) + \frac{\alpha_s}{4\pi} \,
C_i^{4G}(\mu_W) \, .
\end{equation}
These shift at the matching scale are resulted in the
modifications of the effective Wilson coefficients,
\begin{eqnarray}
    \widetilde C_7^{\eff}&=&\left (1+\frac{\alpha_s(\mu)}{\pi}
    \omega_7 (\hat{s})\right ) ( A_7 + A_{77} \; C_7^{4G}(\mu_W) + A_{78} \;
     C_8^{4G}(\mu_W) ) \nonumber \\
    && -\frac{\alpha_{s}(\mu)}{4\pi}\left(C_1^{(0)} F_1^{(7)}(\hat{s})+
    C_2^{(0)} F_2^{(7)}(\hat{s})
+A_8^{(0)} F_8^{(7)}(\hat{s}) +A_{88}^{(0)} \, C_8^{4G}(\mu_W) \,
F_8^{(7)}(\hat{s}) \right) \, ,
    \label{effcoeff7new}  \\
    \widetilde C_9^{\eff}&=&\left (1+\frac{\alpha_s(\mu)}{\pi}
    \omega_9 (\hat{s})\right )
    \left (A_9 + T_9 \, h (\hat m_c^2,
    \hat{s})+U_9 \, h (1,\hat{s}) +
    W_9 \, h (0,\hat{s})  + C_9^{4G}(\mu_W) \right) \nonumber \\
    && -\frac{\alpha_{s}(\mu)}{4\pi}\left(C_1^{(0)} F_1^{(9)}(\hat{s})+
    C_2^{(0)} F_2^{(9)}(\hat{s})+
    A_8^{(0)} F_8^{(9)}(\hat{s})+
    A_{88}^{(0)} \, C_8^{4G}(\mu_W) \, F_8^{(9)}(\hat{s})\right) \, ,
    \label{effcoeff9new} \\
    \widetilde C_{10}^{\eff}&=& \left (1+
    \frac{\alpha_s(\mu)}{\pi}
    \omega_9 (\hat{s})\right ) (A_{10} + C_{10}^{4G} ) \, .
   \label{effcoeff10new}
\end{eqnarray}
The numerical values for the parameters $A_{77}$, $A_{78}$,
$A_{88}^{(0)}$, which incorporate the effects from the running,
can be found  in the same reference \cite{Ali:2002jg}, for the
functions $h (\hat m_c^2,\hat{s})$ and $\omega_9(\hat{s})$, they
are given in Ref.~\cite{BMU}, while $\omega_7(\hat{s})$ and
$F_{1,2,8}^{(7,9)}(\hat{s})$ can be seen in Ref.~\cite{AAGW}.
\begin{figure}[htb]
\begin{center}
\vspace{0.5cm}
    \includegraphics[height=5cm,width=7.5cm]{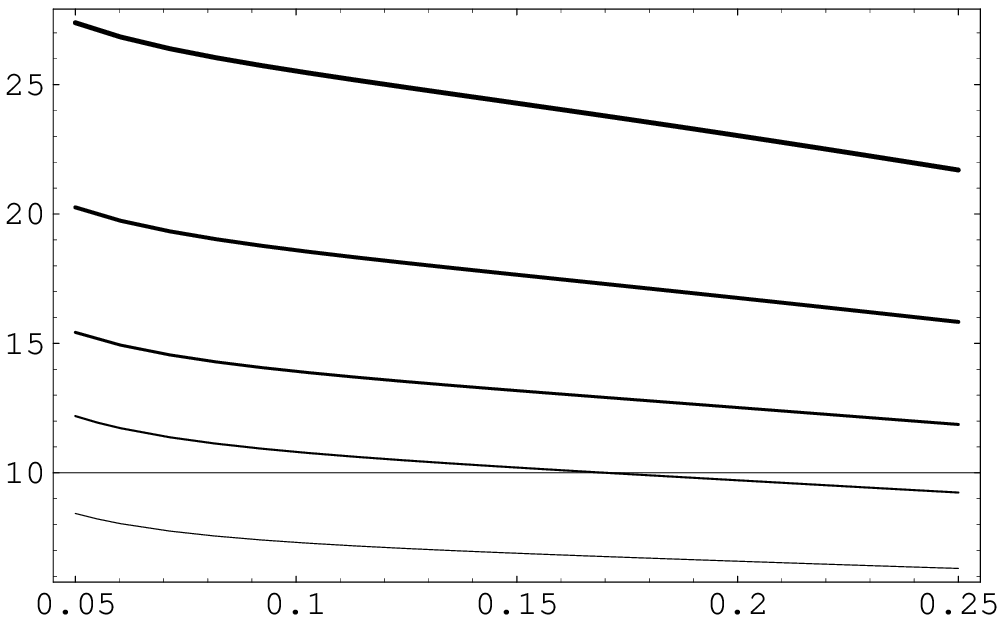}
    \includegraphics[height=5cm,width=7.5cm]{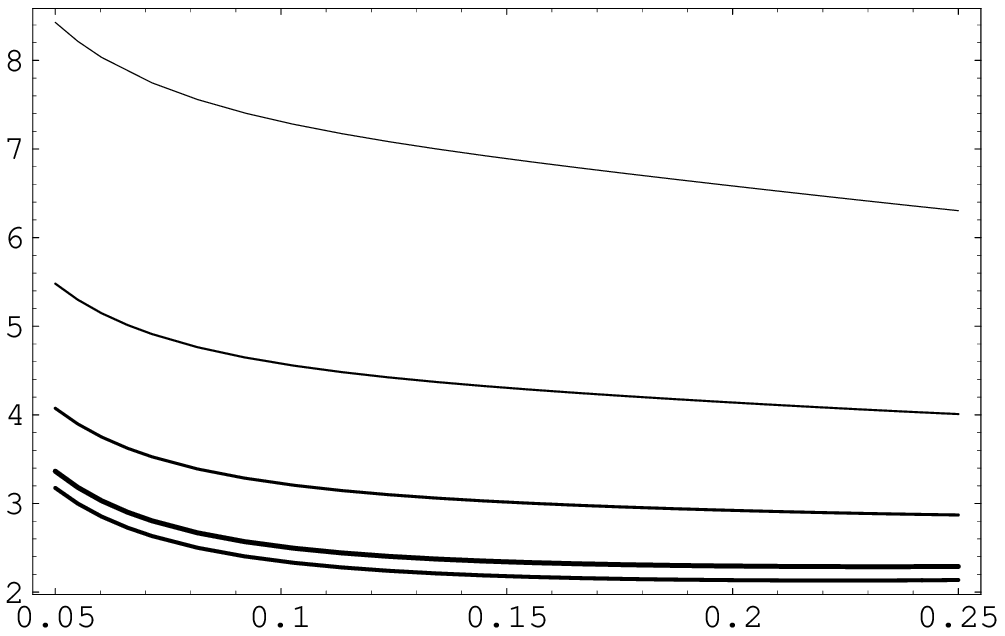}
    \vspace{0cm}
    \caption[]{Branching ratio $\branch^{B\to X_s \ell^+
\ell^-}$ ~$[10^{-6}]$~as a function of $\hat s \in
[0.05,0.25]$(see Eq.(\ref{diff})).
     The four thick lines show the NNLL prediction for $m_{t\prime}=200,300,400$ and $500$
with increasing thickness respectively and the  SM prediction is
the thin line. The figures are obtained at the scale $\mu=5.0$
$GeV$. For the figure at the Left: $\lambda_{t\prime}=-10^{-2}$,
Right: $\lambda_{t\prime}=10^{-2}$.}
    \label{fig1}
    \end{center}
\end{figure}
\begin{figure}[htb]
\begin{center}
\vspace{0.5cm}
    \includegraphics[height=5cm,width=7.5cm]{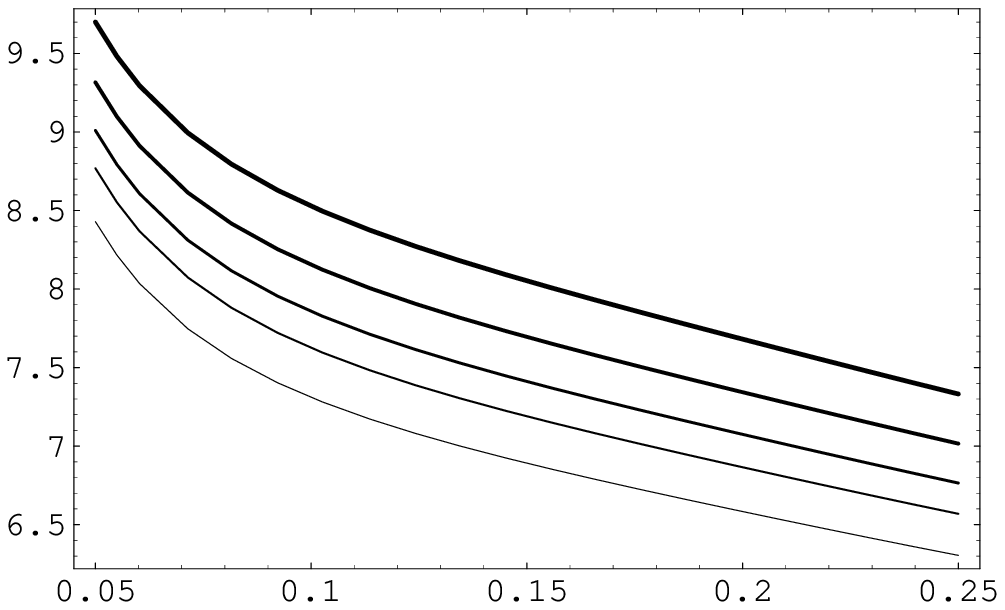}
    \includegraphics[height=5cm,width=7.5cm]{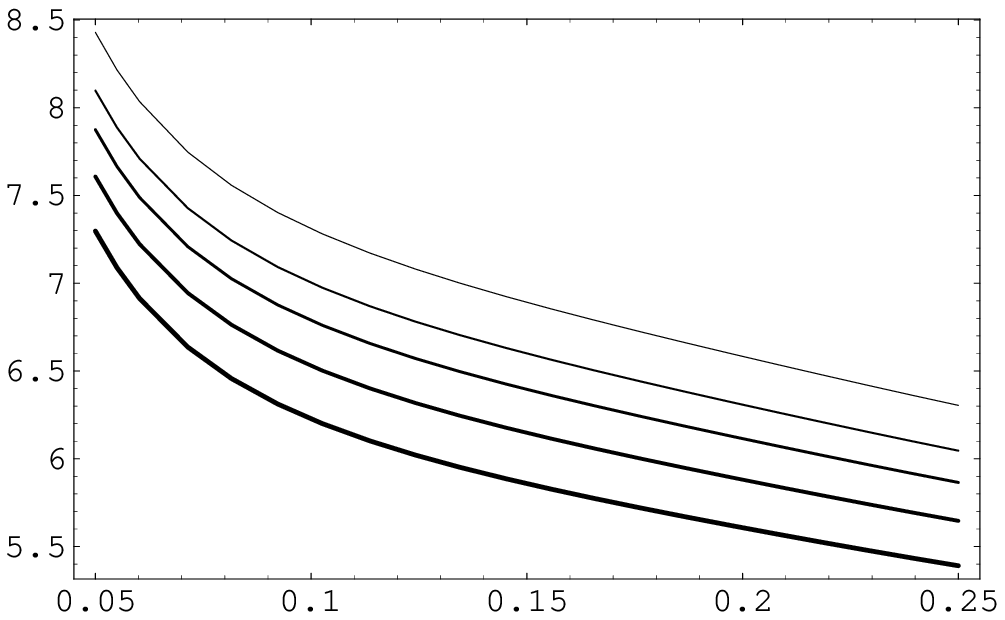}
    \vspace{0cm}
    \caption[]{The same as Fig.\ref{fig1} with the choices, For the figure at the Left:
    $\lambda_{t\prime}=-10^{-3}$,
Right: $\lambda_{t\prime}=10^{-3}$}
    \label{fig2}
    \end{center}
\end{figure}
In order to remove the large uncertainty coming from $m_b$ terms
it is customary to use the following expression \cite{Ali:2002jg}

\begin{equation}
 \branch^{B\to X_s \ell^+
\ell^-} (\hat s)= {\branch_{\hbox{\tiny exp}}^{B \to X_c
e\bar{\nu}} \over \Gamma (B \to X_c e\bar{\nu})} {d\Gamma(B\to
X_s \ell^+ \ell^-) \over d\hat s} \, , \label{diff}
\end{equation}
which can be called as branching ratio. The explicit expression
for the semi-leptonic decay width can be found in Ref. \cite{BMU}.
The branching ratio  with 4G is presented in Figs.
(\ref{fig1},\ref{fig2}) for the choice of the scale $\mu=5$~$GeV$.

 In the figures related with dilepton invariant mass distribution we used the low region $\hat s
\in [0.05,0.25]$ where peaks stemming from $c\bar{c}$ resonances
are expected to be small. During the calculations we take
$\branch_{\hbox{\tiny exp}}^{B \to X_c e\bar{\nu}}=0.1045$.

\subsection{Forward-Backward asymmetry}

We  investigate both, the so-called normalized and the
unnormalized forward-backward asymmetry with 4G model. The double
differential decay width $d^2\Gamma(b \to X_s \ell^+
\ell^-)/(d\sh \, dz)$, ($z=\cos(\theta)$) is expressed as
\cite{Asatrian:2002va}

\bea
\label{doublewidth} \frac{d^2\Gamma(b\to X_s\,
\ell^+\ell^-)}{d\sh \, dz} = &&
    \left(\frac{\alpha_{em}}{4\,\pi}\right)^2
    \frac{G_F^2\, m_{b,\pole}^5\left|V_{ts}^*V_{tb}\right|^2}
    {48\,\pi^3}(1-\sh)^2 \nonumber \\
    &&
    \times \left\{ \frac{3}{4} [(1-z^2)+\sh (1+z^2)] \,
    \left( \left |\widetilde C_9^{\eff}\right |^2 +
    \left |\widetilde C_{10}^{\eff}\right |^2 \right) \,
    \left( 1+\frac{2 \as}{\pi} \, f_{99}(\sh,z) \right) \right. \nn  \\
    && +
    \frac{3}{\sh} [(1+z^2)+\sh(1-z^2)]
    \, \left | \widetilde C_7^{\eff}\right |^2
    \left( 1+\frac{2 \as}{\pi} \, f_{77}(\sh,z) \right)  \nn  \\
    && - 3 \, \sh \, z \, \mbox{Re}(\widetilde C_9^{\eff} \widetilde
    C_{10}^{\eff *})
    \,    \left( 1+\frac{2 \as}{\pi} \, f_{910}(\sh) \right)  \nn  \\
    && + 6  \, \mbox{Re}(\widetilde C_7^{\eff} \widetilde C_9^{\eff *})
    \,    \left( 1+\frac{2 \as}{\pi} \, f_{79}(\sh,z) \right)  \nn  \\
   && \left. - 6 \, z \, \mbox{Re}(\widetilde C_7^{\eff}
    \widetilde C_{10}^{\eff *})
    \,    \left( 1+\frac{2 \as}{\pi} \, f_{710}(\sh) \right) \right\} \, .
\eea

 where $\theta$ is the angle between the momenta of the b quark and the  $ \ell^+$,
 measured in the rest frame of the lepton pair.
  The functions $f_{99}(\sh,z)$, $f_{77}(\sh,z)$,
$f_{910}(\sh)$, $f_{79}(\sh,z)$ and $f_{710}(\sh)$ are the
analogues of $\omega_{99}(\sh)$, $\omega_{77}(\sh)$ and
$\omega_{79}(\sh)$ which can be found in the same reference
\cite{Asatrian:2002va}

\begin{figure}[htb]
\begin{center}
\vspace{0.5cm}
    \includegraphics[height=5cm,width=7cm]{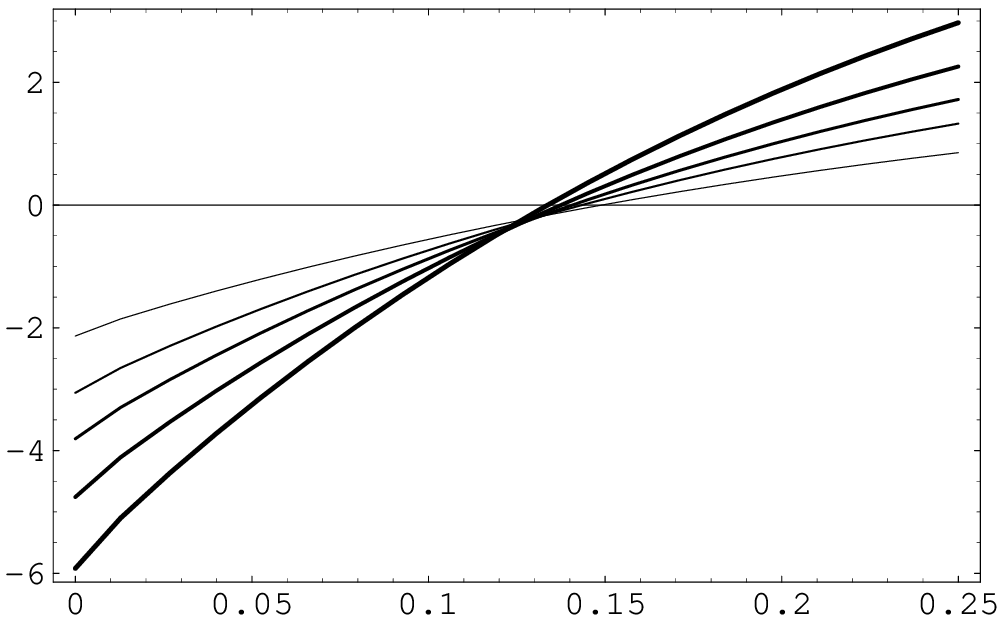}
    \includegraphics[height=5cm,width=7cm]{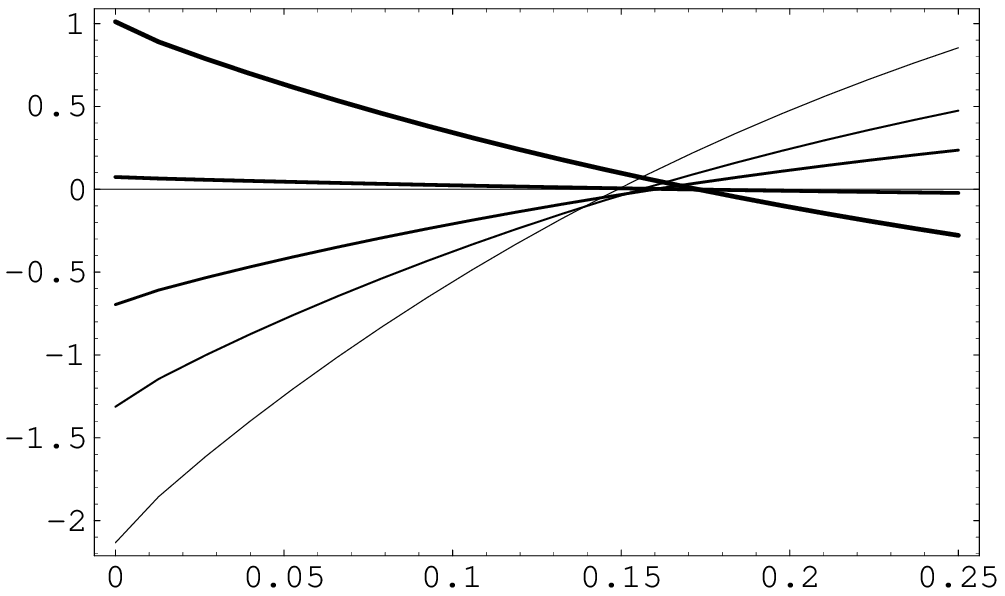}

    \vspace{0cm}
    \caption[]{Unnormalized forward-backward asymmetry
    $A_{\text{FB}}$ ~$[10^{-6}]$~ as a function of $\hat s \in [0,0.25]$ (see
Eq.(\ref{afb})).
     The four thick lines show the NNLL prediction for $m_{t\prime}=200,300,400$ and $500$
with increasing thickness respectively and the  SM prediction is
the thin line. The figures are obtained at the scale $\mu=5.0$
$GeV$. For the figure at the Left: $\lambda_{t\prime}=-10^{-2}$,
Right: $\lambda_{t\prime}=10^{-2}$.}
    \label{fig3}
    \end{center}
\end{figure}

\begin{figure}[htb]
\begin{center}
\vspace{0.5cm}

    \includegraphics[height=5cm,width=7cm]{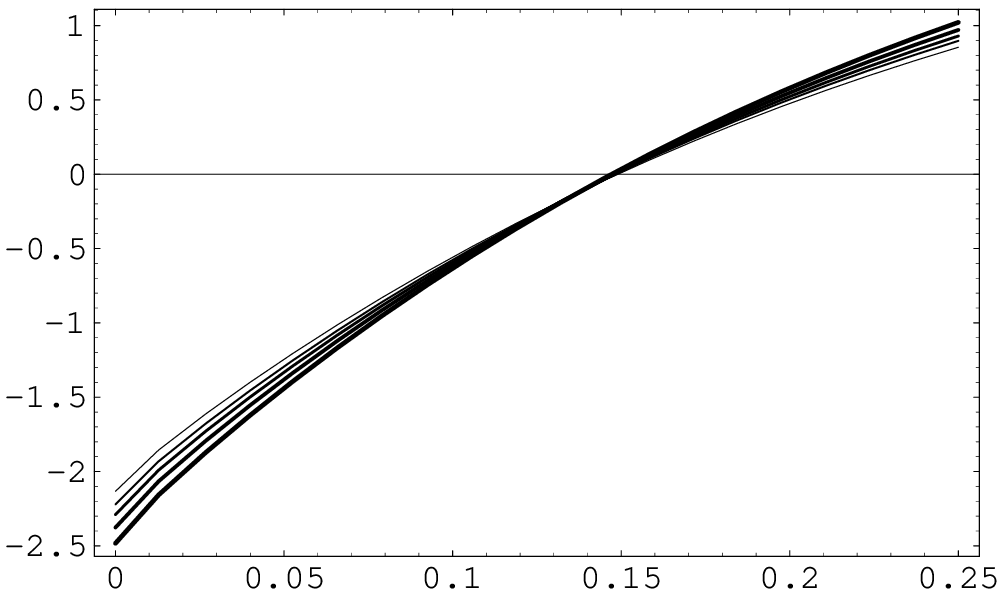}
    \includegraphics[height=5cm,width=7cm]{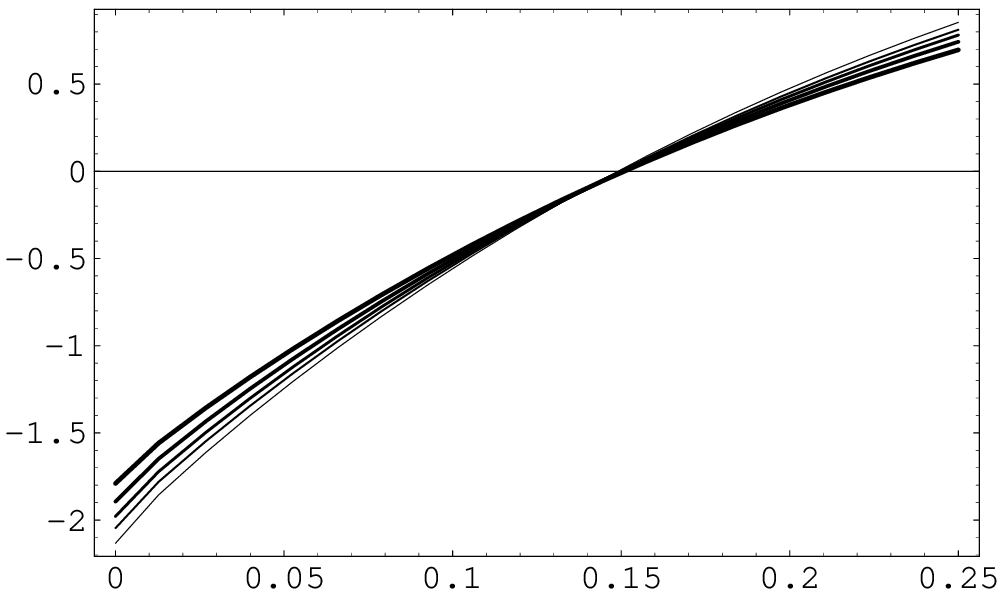}
    \vspace{0cm}
    \caption[]{The same as Fig.\ref{fig3} with the choices: Left:
    $\lambda_{t\prime}=-10^{-3}$,
Right: $\lambda_{t\prime}=10^{-3}$}
    \label{fig4}
    \end{center}
\end{figure}

\begin{figure}[htb]
\begin{center}
\vspace{0.5cm}
    \includegraphics[height=5cm,width=7cm]{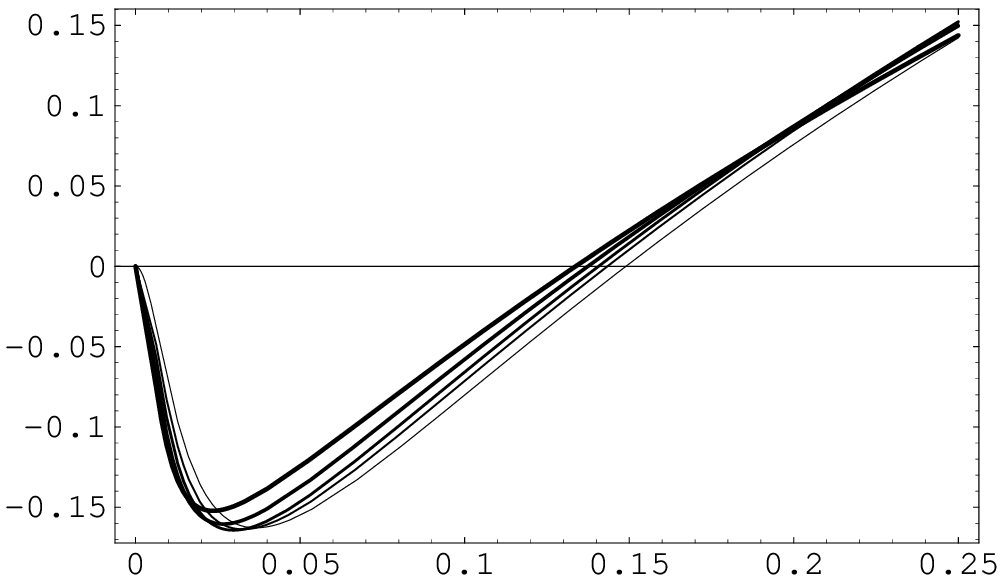}
    \includegraphics[height=5cm,width=7cm]{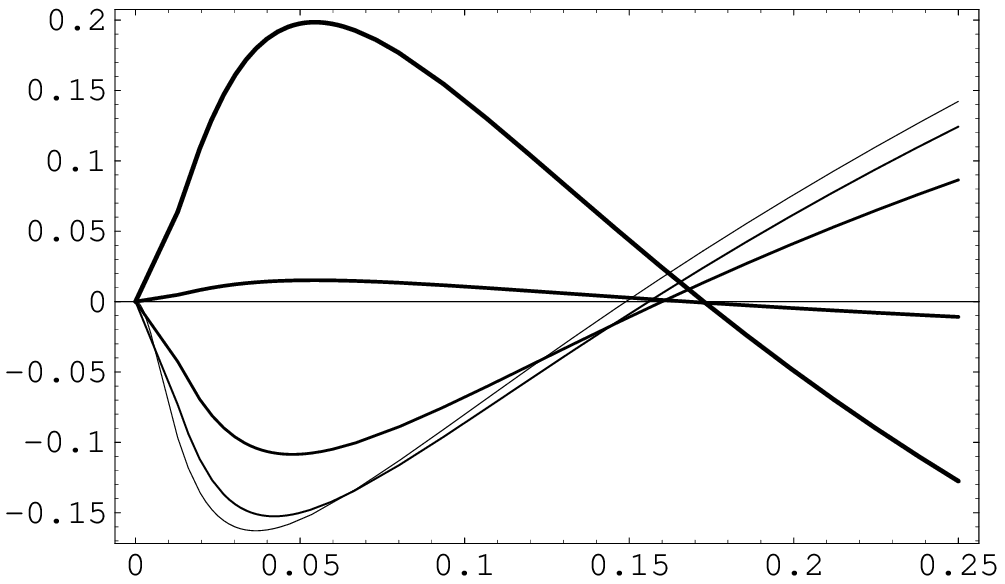}

    \vspace{0cm}
    \caption[]{Normalized forward-backward asymmetry
    $\overline{A}_{\text{FB}}$~$[10^{-6}]$~
as a function of $\hat s \in [0,0.25]$ (see Eq.(\ref{nafb})).
     The four thick lines show the NNLL prediction for $m_{t\prime}=200,300,400$ and $500$
with increasing thickness respectively and the  SM prediction is
the thin line. The figures are obtained at the scale $\mu=5.0$
$GeV$. For the figure at the Left: $\lambda_{t\prime}=-10^{-2}$,
Right: $\lambda_{t\prime}=10^{-2}$.}
        \label{fig5}
    \end{center}
\end{figure}

\begin{figure}[htb]
\begin{center}
\vspace{0.5cm}

    \includegraphics[height=5cm,width=7cm]{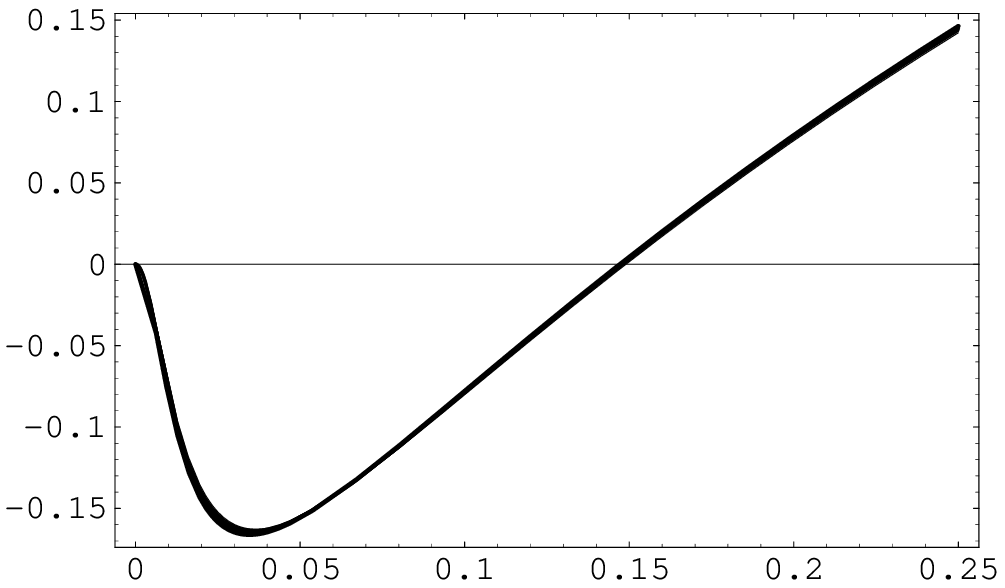}
    \includegraphics[height=5cm,width=7cm]{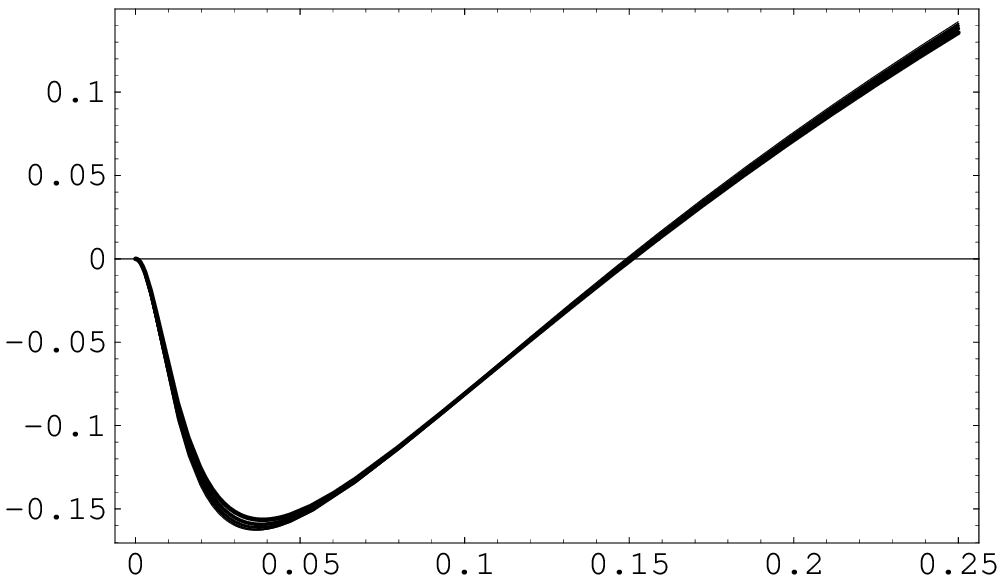}
    \vspace{0cm}
    \caption[]{{The same as Fig.\ref{fig5} with the choices: Left:
    $\lambda_{t\prime}=-10^{-3}$,
Right: $\lambda_{t\prime}=10^{-3}$}.}
    \label{fig6}
    \end{center}
\end{figure}

 The
unnormalized version of forward-backward asymmetry,
$A_{\text{FB}}(\sh)$ is defined as \bea \label{afb}
   A_{\text{FB}}(\sh) &=& \frac{
    \int_{-1}^1 \frac{d^2\Gamma(b\to X_s\, \ell^+\ell^-)}{d\sh \, dz}
    \, \mbox{sgn}(z) \, dz}{\Gamma(B \to X_c e \bar{\nu}_e)} \,
    \branch_{\hbox{\tiny exp}}^{B \to X_c e\bar{\nu}}  \, ,
\eea while the definition of the normalized forward-backward
asymmetry $\overline{A}_{\text{FB}}(\sh)$ reads\bea \label{nafb}
   \overline{A}_{\text{FB}}(\sh) &=& \frac{
    \int_{-1}^1 \frac{d^2\Gamma(b\to X_s\, \ell^+\ell^-)}{d\sh \, dz}
    \, \mbox{sgn}(z) \, dz}{
    \int_{-1}^1 \frac{d^2\Gamma(b\to X_s\, \ell^+\ell^-)}{d\sh \, dz}
     \, dz} \, .
\eea

The position of the zero of the $A_{FB}(\sh_0)=0$ is very
sensitive to 4G effects as it is seen in the figures
(\ref{fig3},\ref{fig5}). However as 4G parameter
$\lambda_{t^\prime}$ decreases expectations of the new model are
getting closer to SM values which can be inferred from
Figs.(\ref{fig4},\ref{fig6})

\subsection{Integrated Branching Ratio}
By suitable  choice of integration limits over $\hat s$ one can
obtain integrated branching ratio in accordance with the
experiment for $e$ and $\mu$, which is already performed, hence we
use the integrated branching ratio expression which has the
following form \cite{Ali:2002jg}:
\bea \label{integ}\branch (B\to X_s \ell^+ \ell^-) &=& 10^{-6}
\times \Big[
       a_1 + a_2 \; |A_7^{\rm tot}|^2 + a_3 \; (|C_9^{\rm 4G}|^2 +
 |C_{10}^{\rm 4G}|^2) \nn \\
&&  + a_4 \; \Re A_7^{\rm tot} \; \Re C_9^{\rm 4G} + a_5 \; \Im
A_7^{\rm tot}
 \; \Im C_9^{\rm 4G}
    + a_6 \;  \Re A_7^{\rm tot} \nn \\
&&  +  a_7 \; \Im  A_7^{\rm tot} + a_8 \; \Re C_9^{\rm 4G} +  a_9
\;
 \Im C_9^{\rm 4G} + a_{10} \; \Re C_{10}^{\rm 4G} \Big]\, ,
\eea
\begin{figure}[htb]
\begin{center}
\vspace{0.5cm}

    \includegraphics[height=5cm,width=7cm]{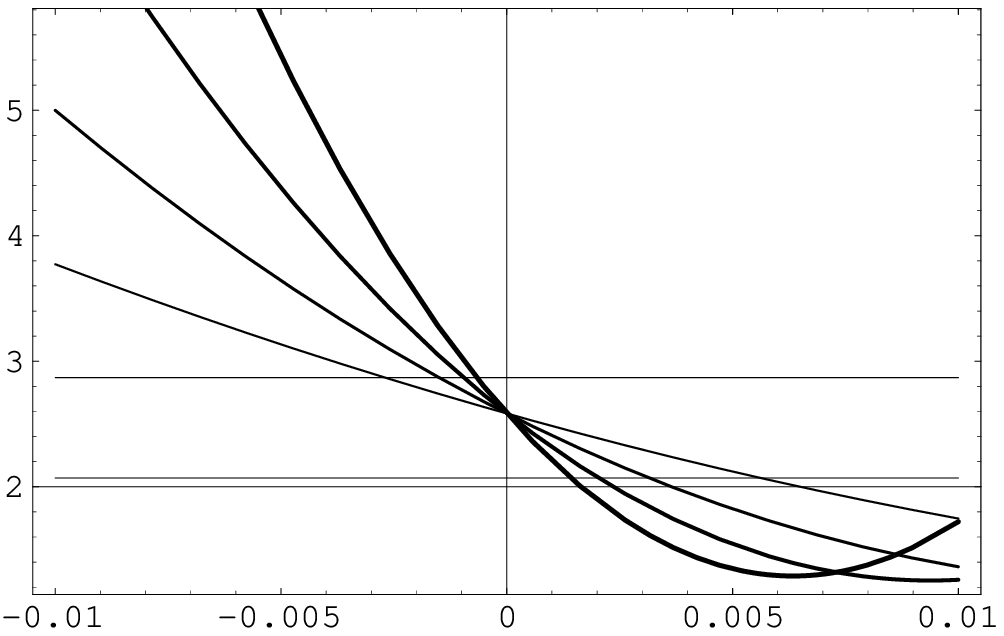}
    \includegraphics[height=5cm,width=7cm]{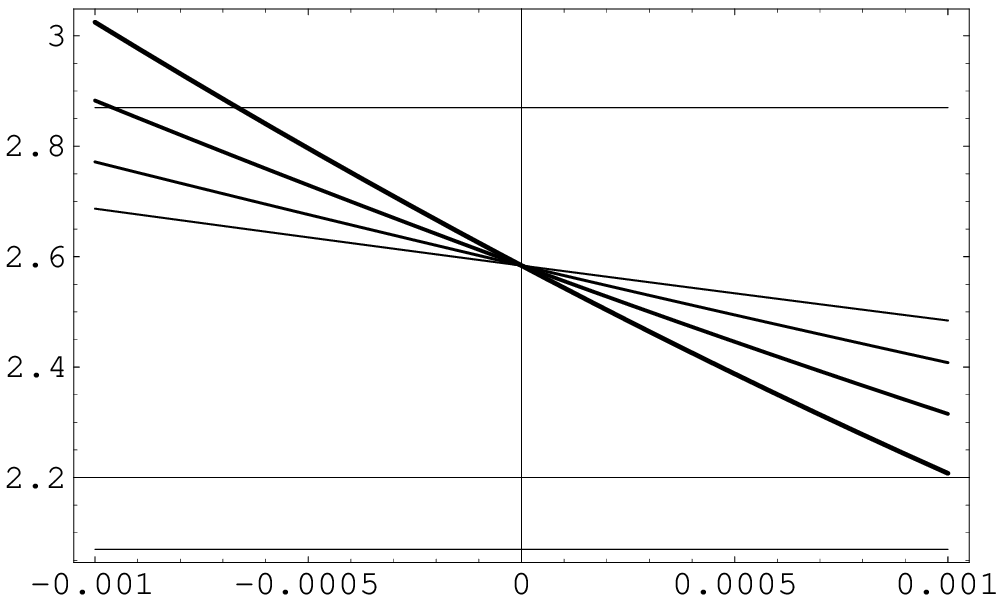}
    \vspace{0cm}
    \caption[]{Integrated Branching ratio \branch $(B\to X_s
    \ell^+\ell^-)$ ~$[10^{-6}]$~as a function of $\lambda_{t\prime}$ for
    $\ell=e$ (see
Eq.(\ref{integ})). In the left figure
    $\lambda_{t\prime}\in[-10^{-2},10^{-2}]$. For the figure at the
    right $\lambda_{t\prime}\in[-10^{-3},10^{-3}]$. In the figures straight lines shows the SM allowed region.}
   \label{fig7}
    \end{center}
\end{figure}

\begin{figure}[htb]
\begin{center}
\vspace{0.5cm}

    \includegraphics[height=5cm,width=7cm]{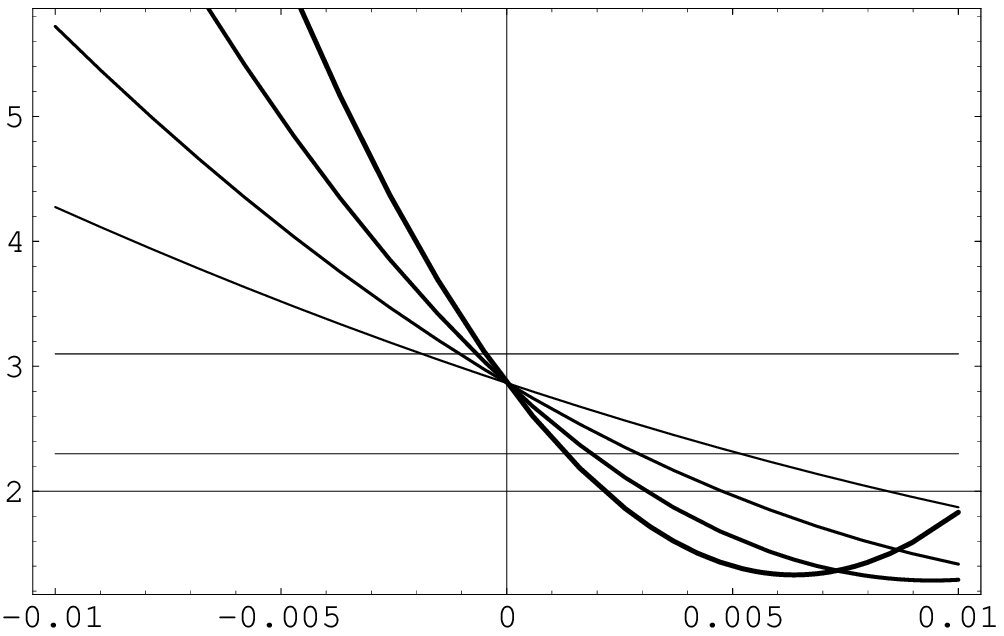}
    \includegraphics[height=5cm,width=7cm]{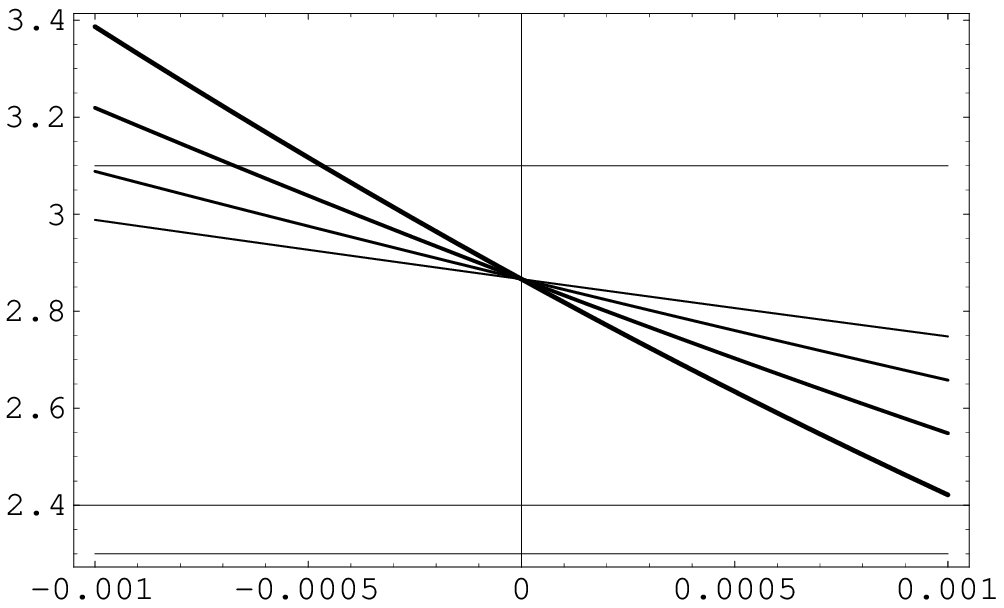}
    \vspace{0cm}
     \caption[]{Integrated Branching ratio \branch $(B\to X_s
    \ell^+\ell^-)$ ~$[10^{-6}]$~as a function of $\lambda_{t\prime}$ for
    $\ell=\mu$. In the left figure
    $\lambda_{t\prime}\in[-10^{-2},10^{-2}]$. For the figure at the
    right $\lambda_{t\prime}\in[-10^{-3},10^{-3}]$. In the figures straight lines show the SM region.}
    \label{fig8}
    \end{center}
\end{figure}
where the numerical value of the coefficients $a_i$ are given in
Table~\ref{table:numcoeff} for $\ell=e,\; \mu$. For the integrated
branching ratios we refer to Figs.(\ref{fig7},\ref{fig8}) of
electron and muon respectively.
\begin{table}[htb]
\begin{center}
\begin{tabular}{|c|c|c|c|c|c|c|c|c|c|c|}\hline
$\ell$&$a_1$&$a_2$&$a_3$&$a_4$&$a_5$&$a_6$&
$a_7$&$a_8$&$a_9$&$a_{10}$ \\ \hline $e$ & 1.9927 & 6.9357 &
0.0640 & 0.5285 & 0.6574 & 0.2673 &
 -0.0586 & 0.4884 & 0.0095 & -0.5288 \\ \hline
$\mu$& 2.3779 & 6.9295 & 0.0753 & 0.6005 & 0.7461 & 0.5955 &
-0.0600 & 0.5828 & 0.0102 & -0.6225 \\ \hline
\end{tabular}
\end{center}
\caption{Numerical values of the coefficients $a_i$ (evaluated at
$\mu_b=5\; \gev$) for the decays $B\to X_s \ell^+ \ell^-$
($\ell=e,\; \mu$), taken from Ref.~\cite{Ali:2002jg}.}
\label{table:numcoeff}
\end{table}

\section{Discussion}

In the sequential fourth generation model, there are basically
two free parameters, mass of new generations and CKM factors
which can have imaginary phases. As a worst scenario, we
decompose $\lambda_{t\prime}= Re[\lambda_{t\prime}]+ I\times
Im[\lambda_{t\prime}]$ and choose the range
$\frac{Im[\lambda_{t\prime}]}{Re[\lambda_{t\prime}]}\leq10^{-2}$;
we checked the effect of this choice and observe that
contribution from the imaginary part can be neglected for all of
the kinematical observables. Naturally, these quantities should
be fixed by respecting experiment. Besides, constraints for CKM
values should be updated by noting that existance of a new
generation can relax the matrix elements of $\text{CKM}_{3\times
3}$, when it is accepted as  a sub-matrix of $\text{CKM}_{4\times
4}$.

Since scale dependency of NNLO calculations of $B\to X_s
\ell^+\ell^-$ are not very  high \cite{Asatrian:2002va}, during
the calculations we set the scale $\mu=5\,~GeV$, use the main
input parameters  as follows,
\begin{eqnarray}
 \alpha_{em}&=&1/133\,, \alpha_{s}(m_Z)=0.119\,,
 G_F = 1.16639 \times 10^{-5} \, {\rm GeV^{-2}} \,,
m_W = 80.33 \, {\rm GeV} \,,\nn\\ m_b &=& 4.8 \, {\rm GeV} \,, m_t
= 176 \, {\rm GeV}, \, m_c = 1.4 \, {\rm GeV} \text{, Wolfenstein
parameters:~}\nn\\ A&=&0.75\,,\lambda=0.221\,,\rho=0.4,\eta=0.2\,.
\end{eqnarray}

 Effects of new physics on kinemaical observables can be summerized as follows:
 \begin{itemize}
 \item Differential decay width $\branch^{B\to X_s \ell^+\ell^-}$
  is presented in figures
Fig.(\ref{fig1},\ref{fig2}), where it is shown that SM prediction
can be strongly enhanced with a new quark for the choice
$\lambda_{t\prime}<0$. It is also possible to supress the decay
width for positive  solutions of $\lambda_{t^\prime}$ which is not
favored.

\item Forward-Backward asymmetry  is also very sensitive to 4G
effects, especially for the choice $\lambda_{t\prime}=10^{-2}$. As
it is seen in Figs.(\ref{fig3},\ref{fig5}), as the mass of
$m_{t^\prime}$ increases it is even possible to have positive
values for $A_{FB}(0)$ which is in contradiction with SM, but
natural in extended models. Once the experimental results related
with this quantity is obtained, it will be a keen test of fourth
generation model. Deviations from the point $\hat s$=0 are
detectable as it is seen in Fig.(\ref{fig4}) for the choice of
$\lambda_{t\prime}\in[-10^{-3},10^{-3}]$, whereas for the same
region we see almost no dependence on the normalized
forward-backward asymmetry in Fig.(\ref{fig6}). While Standard
Model states the central value
$A_{\text{FB}}^{\text{NNLO}}(0)=-(2.30\pm 0.10)\times 10^{-6}$, 4G
predictions cover the range $A_{\text{FB}}^{\text{4G,NNLO}}(0) \in
[-6,1]\times 10^{-6}$\text for the choices
$\lambda_{t^\prime}=-10^{-2}, 10^{-2}$ respectively. For the point
where forward-backward asymmetry vanishes Standard Model result is
$\hat s_0^{\text{NNLO}}=0.162\pm 0.002$ however  4G predictions
are  roughly  $\hat s_0^{\text{4G,NNLO}} \in [0.13,0.18]$.

\item Integrated branching ratios  Figs.(\ref{fig7},\ref{fig8})
strongly depends on the new physics parameters $\lambda_{t\prime}$
and $m_{t^{\prime}}$, therefore it is possible to restrict them by
respecting
 experiments. As it can be deduced from the figures when 4G
 effects are switched off  our calculations are lying on the SM
 ground within error bars \cite{Ali:2002jg}. Similar to branching
 ratio for integrated branching ratios enhancement comes from
 negative choices of $\lambda_{t^\prime}$ which favors smaller values for  $A_{\text{FB}}^{\text{SM,NNLO}}(0)=-(2.30\pm 0.10)\times 10^{-6}$.
\end{itemize}

 To summarize, in this work we
present the predictions of the sequential fourth generation model
for experimentally measurable quantities related with \bsll~decay
which is expected to emerge in the near future thanks to running B
factories. These predictions differ from SM in certain regions,
hence can be used, to differentiate the existence of the fourth
family or to put stringent constrains on the free parameters of
the model, if it exists.

\end{document}